\documentclass[preprint,12pt]{elsarticle}

\usepackage{amsmath}
\usepackage{graphicx}
\usepackage{hyperref}
\usepackage{cleveref}
\usepackage{lmodern}
\usepackage{microtype}
\usepackage[compat=1.1.0]{tikz-feynman}
\usepackage[a4paper,margin=1in]{geometry}
\usepackage{titlesec}

\titlespacing*{\section}{0pt}{*3}{*1}
\titlespacing*{\subsection}{0pt}{*2}{*1}

\biboptions{sort&compress}

\newcommand{\sigmaeff}{\sigma_\mathrm{eff}}
\newcommand{\tvec}[1]{\mathbf{#1}}
\newcommand{\pdfOneDim}{p}
\newcommand{\overlap}{\mathcal{O}}
\newcommand{\fourier}[1]{\tilde{#1}}
\newcommand{\given}{\,|\,}
\newcommand{\eqand}{\quad\text{and}\quad}
\newcommand{\rawrms}[1]{{\scriptstyle\sqrt{\langle #1^2\rangle}}}
\newcommand{\rms}[1]{\vcenter{\hbox{$\rawrms{#1}$}}}

\newcommand{\mb}{\,\text{mb}}
\newcommand{\fm}{\,\text{fm}}
\newcommand{\gev}{\,\text{GeV}}
\newcommand{\tev}{\,\text{TeV}}

\newcommand{\red}[1]{{#1}}

\hypersetup{
    colorlinks=true,
    linkcolor=blue,
    citecolor=blue,
    }

\begin{document}

\journal{Physics Letters B}

\title{Doubling down on proton structure: what double parton scattering can and cannot tell us about partonic separation}

\begin{frontmatter}
\author[xjtlu]{Andrew Fowlie}
\ead{andrew.fowlie@xjtlu.edu.cn}
\author[westlake]{Arthur Moraes}
\ead{arthur@westlake.edu.cn}
\author[xjtlu,uol]{Hao Yang}
\ead{hao.yang@liverpool.ac.uk}

\affiliation[xjtlu]{organization={X-HEP Laboratory},
addressline={Department of Physics, School of Mathematics and Physics, Xi'an Jiaotong-Liverpool University, 111 Ren'ai Road, Suzhou Dushu Lake, Science and Education Innovation District, Suzhou Industrial Park},
postcode={215123},
postcodesep={},
city={Suzhou},
country={China}}

\affiliation[uol]{organization={University of Liverpool},
addressline={Brownlow Hill},
postcode={L69 7ZX},
postcodesep={},
city={Liverpool},
country={United Kingdom}}

\affiliation[westlake]{organization={Westlake University},
addressline={Department of Physics, School of Science, Yungu Campus, 600 Dunyu Road, Xihu District},
postcode={310030},
postcodesep={},
city={Hangzhou},
country={China}}

\begin{abstract}
\red{The} effective cross section $\sigmaeff$ \red{in double parton scattering has been used to construct both lower and upper bounds on the transverse separation between partons in the proton. We critically re-examine these bounds in the factorized framework in which they were proposed,} without assuming any specific parametric form for the \red{proton's transverse profile}. We derive \red{the} sharp lower bound on the transverse partonic separation $\langle r^2\rangle \ge 4\sigmaeff/(9\pi)$, \red{strengthening the previously reported bound $\sigmaeff/(3\pi)$ by a factor of $4/3$. Explicit counterexamples show that no finite upper bound follows from $\sigmaeff$ alone within this framework, even for smooth profiles that are monotonically decreasing on the transverse plane and that decay rapidly in momentum space. Although simplified, this framework clarifies the geometric information carried by the effective cross section and shows that a concentrated component can coexist with substantial probability at large separations. Our results therefore correct the inference that an effective cross section measurement necessarily restricts the partonic separation to a finite interval in this description.}
\end{abstract}

\end{frontmatter}

\section{Introduction}\label{sec:intro}

High-energy hadron collisions, such as proton-proton collisions, are usually described in terms of single parton scattering (SPS), where one parton from each proton participates in a hard scattering. At sufficiently high energies, however, multiple pairs of partons may scatter in the same proton-proton collision. Specifically, two parton scatterings give rise to \emph{double} parton scattering~(DPS)~\cite{Bartalini:2011jp,Diehl:2011yj,Manohar:2012pe,Kasemets:2017vyh,Bartalini:2018abc}. DPS is not only an important background to many Standard Model and new-physics processes~\cite{CDF:2001onq,CMS:2021wlt,ATLAS:2023bft,CMS:2025fpt}, but also provides a unique probe of the spatial transverse structure of the proton beyond that accessible through ordinary SPS processes. 

Experimental studies of DPS processes are commonly characterized by the DPS effective cross section, $\sigmaeff$, extracted through the DPS pocket formula~\cite{CMS:2026evu,ATLAS:2025bcb,ATLAS:2016rnd,CMS:2021lxi,CMS:2022pio,LHCb:2023qgu,LHCb:2023ybt,AxialFieldSpectrometer:1986dfj,CMS:2021qsn,ALICE:2023lsn,LHCb:2016wuo,CMS:2016liw,ATLAS:2016ydt,LHCb:2015wvu,D0:2015rpo,D0:2015dyx,ATLAS:2014ofp,D0:2014owy,D0:2014vql,CMS:2013huw,CMS:2013slh,ATLAS:2014yjd,ATLAS:2013aph,LHCb:2012aiv,D0:2009apj,CDF:1997lmq,CDF:1993sbj,UA2:1991apc}. As one of the most important observables in DPS processes, what information about the proton structure can be inferred from $\sigmaeff$ is an important phenomenological question. Since the idea of the effective cross section $\sigmaeff$ was proposed~\cite{Paver:1982yp}, it has been studied for decades~\cite{Calucci:1997uw,Treleani:2007gi,Rinaldi:2015cya,Bartalini:2018qje,Blok:2010ge,Blok:2011bu,Blok:2013bpa,Blok:2026kxx} and been argued to be closely related to the proton structure~\cite{Calucci:1999yz,Frankfurt:2004kn,Corke:2011yy}. 

\red{In this work, we critically re-examine} upper and lower bounds \red{derived for the transverse} separation between partons in the proton in terms of \red{the effective cross section under the so-called pocket formula framework~\cite{Rinaldi:2018bsf,Rinaldi:2018slz,Rinaldi:2020xtg}. We retain the pocket formula framework because we wish to critique and clarify these previous results found in it, and because it serves as a toy model for understanding the relationship between DPS cross sections and partonic separation. Our goal is not to propose an improved treatment of DPS modelling. We show that no upper bound holds in general, even under the assumptions considered in previous works, and that the lower bound $\sigmaeff/(3\pi)$ reported in Ref.~\cite{Rinaldi:2018slz} can be improved. More general DPS descriptions include longitudinal and transverse correlations, flavour dependence and perturbative parton splitting~\cite{Diehl:2011yj,Blok:2011bu,Blok:2013bpa,Diehl:2017wew}. Some of these effects were also considered in Ref.~\cite{Rinaldi:2018bsf}. Here we focus on the bounds derived under the factorized ansatz to establish precisely what can be inferred about partonic separation from the effective cross section.}

After reviewing models of DPS processes in \cref{sec:effective_cross_section} and introducing definitions, conventions and other preliminaries in \cref{sec:prelim}, in \cref{sec:lower} we derive a lower bound on the expected partonic separation and the root-mean-square (RMS) separation that holds under DPS pocket formula assumptions. Despite its generality, this lower bound improves on previously reported bounds~\cite{Rinaldi:2018bsf,Rinaldi:2018slz,Rinaldi:2020xtg} and we show that it is the sharpest bound possible under our assumptions. Then, in \cref{sec:upper}, we identify the assumptions that would be required to establish an upper bound on the partonic separation. We demonstrate, through explicit examples and analysis, that previously reported upper bounds do not hold in general, and that it would be challenging to construct a robust upper bound, as DPS measurements and interpartonic separations are controlled by different facets of the matter distributions. Lastly, we conclude in \cref{sec:concs}.

\begin{figure}[t]
    \centering
\def\spanx{3.5}
\def\spany{3}

\def\sep{1mm}

\def\col{blue}
\def\altcol{red}
\def\gluoncol{purple}

\newcommand{\addprotons}{
    \def\offset{0.07}
    
    \draw[plain] (0,\spany+\offset+\offset) -- (\spanx,\spany+\offset+\offset);
    \draw[plain] (0,\spany+\offset) -- (\spanx,\spany+\offset);
    \draw[plain] (0,\spany) -- (\spanx,\spany);

    \draw[plain] (0,0) -- (\spanx,0);
    \draw[plain] (0,0-\offset) -- (\spanx,0-\offset);
    \draw[plain] (0,0-\offset-\offset) -- (\spanx,0-\offset-\offset);
}

\pgfmathsetmacro{\halfspanx}{0.5*\spanx}
\pgfmathsetmacro{\quarterspanx}{0.25*\spanx}
\pgfmathsetmacro{\halfspany}{0.5*\spany}
\pgfmathsetmacro{\quarterspany}{0.25*\spany}
\pgfmathsetmacro{\threequarterspany}{0.75*\spany}

\begin{tikzpicture}
  \begin{feynman}  

    \addprotons

    \pgfmathsetmacro{\incomingdx}{0.35*\spanx}
    \pgfmathsetmacro{\outgoingdx}{0.7*0.35*\spanx}
    \pgfmathsetmacro{\outgoingdy}{0.7*\halfspany}

    \vertex[blob, draw=\col, pattern color=\col] (c) at (\halfspanx,\halfspany) {};

    \draw[plain, \col] (\halfspanx-\incomingdx,\spany) -- (c);
    \draw[plain, \col] (\halfspanx-\incomingdx,0) -- (c);

    \draw[plain, \col] (\halfspanx+\outgoingdx,\halfspany+\outgoingdy) -- (c);
    \draw[plain, \col] (\halfspanx+\outgoingdx,\halfspany-\outgoingdy) -- (c);
  \end{feynman}
\end{tikzpicture}
\hspace{\sep}
\begin{tikzpicture}
  \begin{feynman} 

    \addprotons

    \pgfmathsetmacro{\incomingdx}{0.35*\spanx}
    \pgfmathsetmacro{\altincomingdx}{0.3*\spanx}
    \pgfmathsetmacro{\outgoingdx}{0.7*0.35*\spanx}
    \pgfmathsetmacro{\outgoingdy}{0.3*\halfspany}

    \draw[plain] (0,\spany+\offset+\offset) -- (\spanx,\spany+\offset+\offset);
    \draw[plain] (0,\spany+\offset) -- (\spanx,\spany+\offset);
    \draw[plain] (0,\spany) -- (\spanx,\spany);

    \vertex[blob, draw=\col, pattern color=\col] (c) at (\halfspanx,\quarterspany) {};
    \vertex[blob, draw=\altcol, pattern color=\altcol] (d) at (\halfspanx,\threequarterspany) {};

    \draw[plain, \col] (\halfspanx-\incomingdx,\spany) -- (c);
    \draw[plain, \col] (\halfspanx-\incomingdx,0) -- (c);

    \draw[plain, \altcol] (\halfspanx-\altincomingdx,\spany) -- (d);
    \draw[plain, \altcol] (\halfspanx-\altincomingdx,0) -- (d);

    \draw[plain, \col] (\halfspanx+\outgoingdx,\quarterspany+\outgoingdy) -- (c);
    \draw[plain, \col] (\halfspanx+\outgoingdx,\quarterspany-\outgoingdy) -- (c);
    
    \draw[plain, \altcol] (\halfspanx+\outgoingdx,\threequarterspany+\outgoingdy) -- (d);
    \draw[plain, \altcol] (\halfspanx+\outgoingdx,\threequarterspany-\outgoingdy) -- (d);
  \end{feynman}
\end{tikzpicture}
    \caption{Schematic Feynman diagrams for SPS (left) and DPS (right) between partons in a collision between two protons.}
    \label{fig:dps}
\end{figure}
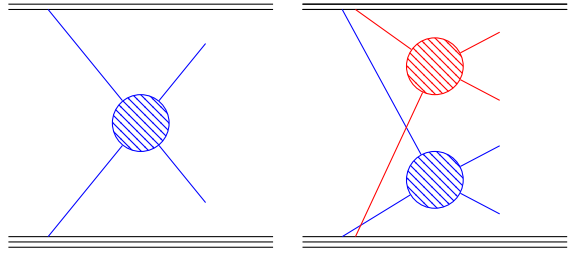

\section{Effective cross section}\label{sec:effective_cross_section}

\red{To allow us to critique and clarify existing results, we use the same framework as Ref.~\cite{Rinaldi:2018slz, Rinaldi:2018bsf,Rinaldi:2020xtg}}. We begin by clarifying the DPS pocket formula and the definition of the effective cross section. We start from the DPS cross section $\sigma_{\mathrm{AB}}^\mathrm{DPS}$ given in the collinear factorization~\cite{Diehl:2011yj},
\begin{equation}\label{eq:collinear_factorization}
    \begin{split}
        \sigma_\mathrm{AB}^\mathrm{DPS}
        = 
        S_{\mathrm{AB}}
        \smash{\sum\limits_{i,j,k,l}\int}
        d^2\tvec{b}
        \, dx_1 dx_2 dx_1^\prime dx_2^\prime 
        \,
        \Big[
        & D_{ij}(x_1,x_2;\tvec{b};Q_1^2,Q_2^2) \times \\
        & 
        D_{kl}(x'_1,x'_2;\tvec{b};Q_1^2,Q_2^2) \times \\
        & \hat{\sigma}_{ik}^{\mathrm{A}}(x_1,x'_1,Q_1^2) \times 
        \hat{\sigma}_{jl}^{\mathrm{B}}(x_2,x'_2,Q_2^2)
        \Big],
    \end{split}
\end{equation} 
where $S_\mathrm{AB}$ is a symmetry factor that equals $1/2$ when the final states $\mathrm{A}$ and $\mathrm{B}$ are identical and one otherwise; $D_{ij}$ are double parton distribution functions~(dPDFs) for partons $i$ and $j$; and $\hat{\sigma}_{ik}^{\mathrm{A}}$ are sub-process parton-level cross sections for the production of final state $\mathrm{A}$ from incoming partons $i$ and $k$. The dPDFs depend on the momentum fractions, $x_1$ and $x_2$, and the \red{hard} scales, $Q_1$ and $Q_2$, of \red{the processes involving} partons $i$ and $j$, respectively, as well as on the transverse separation $\tvec{b}$ between partons $i$ and $j$, or equivalently, the transverse distance between two partonic scatterings. Under common assumptions\red{~\cite{Rinaldi:2015cya,Rinaldi:2018bsf,Rinaldi:2018slz,Rinaldi:2020xtg}}, the dPDFs factorize as
\begin{equation}
    D_{ij}(x_1,x_2;\tvec{b};Q_1^2,Q_2^2) \simeq f_i(x_1;Q_1^2) \, f_j(x_2;Q_2^2) \, \overlap(\tvec{b})
    \label{eq:dps_factorization}
\end{equation}
where $f_i(x;Q^2)$ are the parton distribution functions of parton $i$ at momentum fraction $x$ and energy scale $Q$; and $\overlap(\tvec{b})$ is called the overlap function. The overlap function $\overlap(\tvec{b})$ must be normalized, i.e., $\int \overlap(\tvec{b}) d^2\tvec{b}=1$. By substituting \cref{eq:dps_factorization} into \cref{eq:collinear_factorization}, we obtain
\begin{equation}
    \begin{split}
        \sigma_{\mathrm{AB}}^\mathrm{DPS} ={}& S_\mathrm{AB}
        \sum_{i,k} \int dx_1 dx_1^\prime \, f_{i}(x_1;Q_1^2) \, f_{k}(x'_1;Q_1^2) \,\hat{\sigma}_{ik}^{\mathrm{A}}(x_1,x'_1,Q_1^2)\\
        &\times
        \sum_{j,l} \int dx_2 dx_2^\prime \, f_{j}(x_2;Q_2^2) \, f_{l}(x'_2;Q_2^2) \, \hat{\sigma}_{jl}^{\mathrm{B}}(x_2,x'_2,Q_2^2)\\
        &\times\int\overlap^2(\tvec{b})d^2\tvec{b}
    \end{split}
\end{equation}
The factorization \cref{eq:dps_factorization} allows us to rewrite the DPS cross section in the form of the DPS pocket formula,
\begin{equation}\label{eq:pocket_formula}
    \sigma_\mathrm{AB}^\mathrm{DPS}=S_\mathrm{AB} \frac{\sigma_\mathrm{A}\sigma_\mathrm{B}}{\sigmaeff}
\end{equation}
where $\sigma_\mathrm{A}$ and $\sigma_\mathrm{B}$ are SPS cross sections for final states $\mathrm{A}$ and $\mathrm{B}$, respectively, e.g.,
\begin{equation}
    \sigma_\mathrm{A} = \sum\limits_{i,k} \int dx_1 dx_1^\prime f_{i}(x_1;Q_1^2) \, f_{k}(x'_1;Q_1^2) \, \hat{\sigma}_{ik}^{\mathrm{A}}(x_1,x'_1,Q_1^2)
\end{equation}
and the effective cross section $\sigmaeff$ is defined as
\begin{equation}
    \sigmaeff^{-1}=\int{\overlap^2(\tvec{b})d^2\tvec{b}}
    \label{eq:def_sigma_eff}
\end{equation}
Within the DPS pocket formula framework, the overlap function $\overlap(\tvec{b})$ tells us the probability density for the transverse separation $\tvec{b}$ between two partons inside the same proton. Thus, it is positive and normalized. The effective cross section provides a measure of the spread of the overlap function. 

We stress that under other frameworks, the overlap function may have a different interpretation. \red{The probabilistic interpretation here relies on the factorization ansatz in \cref{eq:dps_factorization}: the same overlap function is assigned to the parton pairs in both protons, independently of their longitudinal momentum fractions and flavours. More general DPS descriptions include longitudinal and transverse correlations, flavour dependence, and perturbative parton splitting~\cite{Diehl:2011yj,Blok:2011bu,Blok:2013bpa,Diehl:2017wew}.  However, we retain} the DPS pocket formula \red{framework because we wish to critique results found in the DPS pocket formula framework and because it serves as a toy model for clarifying and understanding the relationship between DPS cross sections and partonic separation.  A restricted extension is possible if the common transverse profile depends on the hard scales while remaining independent of momentum fractions and flavours,  
\begin{equation}\label{eq:relaxed_pocket}
    D_{ij}(x_1,x_2;\tvec{b};Q_1^2,Q_2^2) \simeq f_i(x_1;Q_1^2) \, f_j(x_2;Q_2^2) \, \overlap(\tvec{b}; Q_1^2, Q_2^2).
\end{equation}
As this retains factorization between partonic separation and the Bjorken-$x$ variables, however, it does not capture the full effects of perturbative parton splitting and other sources of correlation. Whilst  our results would hold under \cref{eq:relaxed_pocket}, the overlap function would differ between processes and there would no longer be a process-universal effective cross section.}

\section{Preliminaries}\label{sec:prelim}

Before analyzing upper and lower bounds on partonic separation, we first introduce preliminary definitions and theorems. First, the Cauchy-Schwarz inequality~\cite{Axler:2024} implies that for two square-integrable, real-valued functions, $f(x)$ and $g(x)$, defined on the domain $S$, 
\begin{equation}\label{eq:cs}
    \int_S f(x)g(x)dx \le \sqrt{\int_S f(x)^2 dx} \, \sqrt{\int_S g(x)^2 dx}.
\end{equation}
We use this to bound integrals of the overlap function. For example, suppose the overlap function $\overlap(\tvec{b})$ has a finite area of support, $A$, which means that outside the area $A$ the overlap function is zero almost everywhere. Then by~\cref{eq:cs}, we have
\begin{equation}
    1=\left(\int_A\overlap(\tvec{b}) d^2\tvec{b}\right)^2\le 
    \left(\int_{A}\overlap^2(\tvec{b})d^2\tvec{b}\right)\left(\int_A1d^2\tvec{b}\right)=\sigmaeff^{-1}A
\end{equation} 
Hence, the supporting area of the overlap function provides an upper bound on the effective cross section, $\sigmaeff \le A$, and the equality holds if and only if the overlap function is uniform over its supporting area. 

Second, we stress that the overlap function $\overlap$ is not a probability density function of the relative distance $r$. In general, we may write the overlap function as a density function in polar coordinates as,
\begin{equation}
    \pdfOneDim(r, \theta) = \overlap(x = r \cos\theta, y = r\sin\theta) r =  \pdfOneDim(r) \, \pdfOneDim(\theta \given r).
\end{equation}
The one-dimensional probability density function of the relative distance $r$, $\pdfOneDim(r)$, may be found by 
\begin{equation}\label{eq:pdf_r_overlap}
    \pdfOneDim(r) = \int \delta\!\left(r - \sqrt{x^2 + y^2}\right)  \overlap(x, y) \, dx dy = \int_0^{2\pi} \overlap(r, \theta) \, r d\theta.
\end{equation}
It may be reasonable to assume that the overlap function is rotationally symmetric around zero separation, $\tvec b = 0$, since there is no evidence of a special direction for the transverse structure of the proton and since the beams are not prepared in any special way, e.g., polarized. Assuming rotational symmetry, \cref{eq:pdf_r_overlap} becomes
\begin{equation}\label{eq:one_dim_symmetric}
    \pdfOneDim(r) = 2 \pi r \mskip1mu \overlap(r)
\end{equation}
Furthermore, we may express the effective cross section in polar coordinates by
\begin{equation}
    \sigmaeff^{-1} = \int \overlap^2(x, y) \, dx dy = \int \frac{\pdfOneDim^2(r)}{r} \, \pdfOneDim^2(\theta \given r) \, dr d\theta.
\end{equation}
Assuming rotational symmetry, $\pdfOneDim(\theta \given r) = 1/(2\pi)$ and thus,
\begin{equation}\label{eq:sigma_eff_symmetric}
\sigmaeff^{-1} = \int \frac{\pdfOneDim^2(r)}{2\pi r} dr.
\end{equation}
Without rotational symmetry the effective cross section depends on both the radial and angular distributions. We can, though, establish a bound:
\begin{equation}\label{eq:separable}
    \sigmaeff^{-1} = \int \frac{\pdfOneDim^2(r)}{r} \, \pdfOneDim^2(\theta \given r) \, dr d\theta = \int \frac{\pdfOneDim^2(r)}{r} \left[\int_0^{2\pi} \pdfOneDim^2(\theta \given r) d\theta\right]  dr\ge \int \frac{\pdfOneDim^2(r)}{2\pi r} dr
\end{equation}
since, by the Cauchy-Schwarz inequality~\cref{eq:cs}, for all $r$,
\begin{equation}
    \int_0^{2\pi} \pdfOneDim^2(\theta \given r) d\theta \ge \frac{1}{2\pi}.
\end{equation}
The equality occurs in the rotationally symmetric case.

Lastly, we introduce the structure function $\fourier{\overlap}(k)$, which is the 2D Fourier transformation of the overlap function. For a rotationally symmetric overlap function, it can also be represented as
\begin{equation}\label{eq:fourier_overlap}
    \fourier{\overlap}(k) \equiv 2\pi\int_0^\infty r \overlap(r) J_0(kr) dr = \int_0^\infty \pdfOneDim(r) J_0(kr) dr,
\end{equation}
where $J_0$ is the Bessel function of the first kind with order $0$ and $k$ denotes a conjugate momentum variable. The normalization of the radial pdf $\pdfOneDim(r)$, the effective cross section, and the RMS of the separation can be written in momentum space~\cite{Rinaldi:2018slz} as, 
\begin{align}
    1 &= \int_0^{\infty} \pdfOneDim(r) dr = \fourier{\overlap}(0), \\
    \sigmaeff^{-1} &= \int \frac{\pdfOneDim^2(r)}{2\pi r} dr = \frac{1}{2\pi}\int_0^\infty k \, \fourier{\overlap}^2(k) dk,\label{eq:k_sigma_eff}\\
    \langle r^2 \rangle &= -2 \left.\frac{d^2\fourier{\overlap}(k)}{dk^2}\right|_{k=0},\label{eq:k_r2}
\end{align}
respectively, where we assumed rotational symmetry to compute the effective cross section and, assuming rotational symmetry, the one-dimensional pdf $\pdfOneDim(r)$ can be found from the overlap function by \cref{eq:one_dim_symmetric}. Because $|J_0(kr)| \le 1$, 
\begin{equation}\label{eq:hankel_upper}
    |\fourier{\overlap}(k)| \le 1,
\end{equation}
by definition \cref{eq:fourier_overlap} and by the fact that $p(r)$ is a pdf.

\section{A lower bound for partonic separation}\label{sec:lower}

Ref.~\cite{Rinaldi:2018slz} reports a lower bound on partonic separation,
\begin{equation}\label{eq:previous_lower_bound}
    \langle r^2 \rangle \ge \frac{\sigmaeff}{3\pi}.
\end{equation}
The existence of such a bound is intuitive: to decrease the inverse effective cross section, one must spread the probability mass of the overlap function to larger separations, which inevitably increases the typical parton separation.

We now prove a stronger lower bound than \cref{eq:previous_lower_bound}, not even assuming rotational symmetry. For any $R>0$ and any pdf $\pdfOneDim(r)$, 
\begin{equation}
    1-\frac{\langle r^2\rangle}{R^2}=\int_{0}^{\infty}\left(1-\frac{r^2}{R^2}\right)\pdfOneDim(r)dr\le \int_{0}^{R}\left(1-\frac{r^2}{R^2}\right)\pdfOneDim(r)dr,
    \label{eq:lemma-inequality}
\end{equation}
where the inequality holds due to the positivity of the pdf $\pdfOneDim(r)$. Applying the Cauchy-Schwarz inequality~\cref{eq:cs} to the right-hand side of \cref{eq:lemma-inequality} and using the inequality \cref{eq:separable}, we obtain
\begin{align}
    \int_{0}^{R}\left(1-\frac{r^2}{R^2}\right)\pdfOneDim(r)dr
    &\le \left(\int_0^R\frac{\pdfOneDim^2(r)}{2\pi r}dr\right)^{1/2} 
    \left( \int_{0}^{R}2\pi r\left(1-\frac{r^2}{R^2}\right)^{2}dr\right)^{1/2} \label{eq:cauchy-shwarz}\\
    &\le \left(\int_0^\infty\frac{\pdfOneDim^2(r)}{2\pi r}dr\right)^{1/2}\label{eq:third-inequality}
    \left( \int_{0}^{R}2\pi r\left(1-\frac{r^2}{R^2}\right)^{2}dr\right)^{1/2}\\
    &\le\sqrt{\frac{\pi}{3}} \frac{R}{\sqrt{\sigmaeff}}
\end{align}
Combining this with \cref{eq:lemma-inequality}, we have
\begin{equation}
    \frac{\langle r^2\rangle}{R^2}+ \sqrt{\frac{\pi}{3}} \frac{R}{\sqrt{\sigmaeff}} \ge 1
\end{equation}
This must hold for any choice of $R > 0$; minimizing the left-hand side with respect to $R$ yields
\begin{equation}\label{eq:new-lower-bound}
   \langle r^2\rangle \ge  \frac{4}{9\pi} \sigmaeff
\end{equation}
This lower bound improves on \cref{eq:previous_lower_bound} from~\cite{Rinaldi:2018slz} by a factor $4/3$ and can be saturated for the rotationally symmetric, non-negative and normalized pdf,
\begin{equation}\label{eq:saturate_lower}
    \pdfOneDim_{*}(r)=
        \begin{cases}
            \frac{4 r \left(R^{2} - r^{2}\right)}{R^{4}} & 0\le r\le R \\
            0 & \text{otherwise}
        \end{cases}
\end{equation}
as it results in
\begin{equation}
    \langle r^2 \rangle = \frac{R^2}{3} \eqand \sigmaeff^{-1} = \frac{4}{3 \pi R^{2}},
\end{equation}
and thus $\langle r^2\rangle = 4\sigmaeff / (9\pi)$. The existence of $\pdfOneDim_{*}(r)$ in \cref{eq:saturate_lower} shows that \cref{eq:new-lower-bound} is sharp, which means that it is the greatest lower bound within the framework introduced in \cref{sec:effective_cross_section} and any attempt to raise this lower bound will require additional assumptions on the transverse profile. \red{The derivation applies equally to the ansatz in \cref{eq:relaxed_pocket} in which the overlap function carries hard scale dependence. The bound therefore applies separately at each fixed pair of hard scales,
\begin{equation}
    \langle r^2\rangle_{Q_1 Q_2} \ge \frac{4}{9\pi} \sigmaeff(Q_1^2,Q_2^2).
\end{equation}
Here, $Q_1$ and $Q_2$ are the hard scales of the two subprocess in the DPS interaction, and the RMS partonic separation applies to partons defined those scales.}

The experimental measurements for $\sigmaeff$ to date range from $5\mb$ to $20\mb$~\cite{CMS:2026evu,ATLAS:2025bcb,ATLAS:2016rnd,CMS:2021lxi,CMS:2022pio,LHCb:2023qgu,LHCb:2023ybt,AxialFieldSpectrometer:1986dfj,CMS:2021qsn,ALICE:2023lsn,LHCb:2016wuo,CMS:2016liw,ATLAS:2016ydt,LHCb:2015wvu,D0:2015rpo,D0:2015dyx,ATLAS:2014ofp,D0:2014owy,D0:2014vql,CMS:2013huw,CMS:2013slh,ATLAS:2014yjd,ATLAS:2013aph,LHCb:2012aiv,D0:2009apj,CDF:1997lmq,CDF:1993sbj,UA2:1991apc}. These measurements use different initial states, such as $pp$ or $p\bar{p}$; use different centre-of-mass energies, from $630\gev$ to $13\tev$; select different final states, such as $\gamma$ + jets, $J\psi J\psi$ and $W^\pm W^\pm$; make different phase space cuts; and use different DPS/SPS separation methods. However, one common assumption is that they all determine $\sigmaeff$ based on \cref{eq:pocket_formula}. Taking $\sigmaeff=15\mb$, as commonly used in theoretical calculations, we find $\rms{r} \ge 0.46\fm$. Though not directly comparable, note that the observed RMS electric and magnetic radii of the proton are $0.8407\pm0.0006\fm$ and $0.851\pm0.026\fm$, respectively~\cite{ParticleDataGroup:2026aaa}. These radii may differ, as the electric and magnetic radii characterize charge and magnetic moment density, respectively, whereas interpartonic separation characterizes transverse structure and is not weighted by parton charge or spin.

Similar reasoning leads to the sharp lower bound for expected separation,
\begin{equation}
    \langle r \rangle \ge \sqrt{\frac{3\sigmaeff}{8\pi}}
\end{equation}
We focus on RMS separation following~\cite{Rinaldi:2015cya}. 

\section{An upper bound for partonic separation}\label{sec:upper}

To compute an upper bound, we assume rotational symmetry of the overlap function and express the problem in momentum space using the Fourier transformation of the overlap function \cref{eq:fourier_overlap}. To reconstruct previous upper bounds on the mean-squared separation, we impose the following piecewise assumptions on the structure function for low $k \le k_l$, intermediate $k_l < k < k_u$, and high $k \ge k_u$: 
\begin{alignat}{2}
    1 - \frac{k^2}{k_l^2} \le \fourier{\overlap}(k) &\le 1 &&\quad\text{ for } k \le k_{l}\label{eq:assumption_low_k}\\
    |\fourier{\overlap}(k)| &\le 1 &&\quad\text{ for } k_l < k < k_u\label{eq:assumption_intermediate_k}\\
    |\fourier{\overlap}(k)| &\le \frac{k_u^2}{k^2} &&\quad\text{ for } k \ge k_u \label{eq:assumption_high_k}
\end{alignat}
The bounds $|\fourier{\overlap}(k)| \le 1$ apply automatically by \cref{eq:hankel_upper}.
These assumptions are more restrictive than those in~\red{Ref.}~\cite{Rinaldi:2018bsf}, which only applied the high $k$ assumption \cref{eq:assumption_high_k}. The requirement that the structure function decays at least as fast as $1/k^2$ in momentum-space suppresses narrow structure in position-space. The argument for it appears to be that partonic structure functions should be no harder than electromagnetic ones, which are soft and show no short-scale structure, i.e., $1/k^2$ drop-off. Although this is a strong and model-dependent assumption, as we shall see it is not enough. We further assume that $\fourier{\overlap}(k)$ is twice differentiable at least at $k = 0$. If it were not, \cref{eq:k_r2} and the mean-squared separation would not exist. As we seek to bound the mean-squared separation, this is thus a natural assumption.

Using these assumptions, we may bound the effective cross section,
\begin{align}\label{eq:sigma_eff_bound}
    \sigmaeff^{-1} &= \frac{1}{2\pi} \int_0^{k_u}  k \, \fourier{\overlap}^2(k) dk + \frac{1}{2\pi} \int_{k_u}^\infty k \, \fourier{\overlap}^2(k) dk \\
    &\le 
    \frac{1}{2\pi} \int_0^{k_u} k dk + \frac{1}{2\pi} \int_{k_u}^\infty \frac{k_u^4}{k^3} dk\\
    &= \frac{1}{2\pi} \frac{k_u^2}{2} + \frac{1}{2\pi} \frac{k_u^2}{2} = \frac{1}{2\pi} k_u^2
\end{align}
The assumption that $\fourier{\overlap}(k)$ was twice differentiable at $k=0$ and \cref{eq:assumption_low_k} imply that
\begin{equation}
    \left.\frac{d^2\fourier{\overlap}(k)}{dk^2}\right|_{k=0} \ge -\frac{2}{k_l^2}
\end{equation}
such that
\begin{equation}\label{eq:r2_bound}
    \langle r^2 \rangle \le \frac{4}{k_l^2}.
\end{equation}
Multiplying \cref{eq:r2_bound,eq:sigma_eff_bound}, we obtain
\begin{equation}\label{eq:r2_bound_weak}
    \sigmaeff^{-1} \langle r^2 \rangle \le \frac{2}{\pi} \frac{k_u^2}{k_l^2}
\end{equation}
This is an upper bound, but it is arbitrarily weak when there is an intermediate region $k_l < k < k_u$ that we do not control. The definitions \cref{eq:k_sigma_eff,eq:k_r2} show the difficulty in establishing a bound between the expected separation squared and the effective cross section: the former depends on the curvature of the structure function at $k = 0$; the latter depends on an integral over the structure function across all $k$.

For example, to make \cref{eq:r2_bound_weak} arbitrarily weak, consider a mixture of two Gaussian structure functions,
\begin{equation}\label{eq:counter_example}
    \fourier{\overlap}(k) = \frac{1}{2} \left(e^{-\frac12a_1 k^2} + e^{-\frac12a_2 k^2}\right)
\end{equation}
with $a_2 \gg a_1$. The $a_1$ Gaussian gives a broad plateau that eventually falls off, setting $k_u$. The $a_2$ Gaussian gives a narrow peak, setting $k_l$. The hierarchy $a_2 \gg a_1$ thus leads to $k_u / k_l \gg 1$ such that \cref{eq:r2_bound_weak} is not useful. Computing the terms explicitly,
\begin{align}
    \langle r^2 \rangle &= a_1+a_2\\
    \sigmaeff &= \frac{16\pi a_1a_2(a_1+a_2)}{(a_1+a_2)^2+4a_1a_2},
\end{align}
we see that we can make $\langle r^2 \rangle$ arbitrarily large by taking $a_2 \to \infty$; at the same time, we can make $\sigmaeff$ arbitrarily small by taking $a_1 a_2 \to 0$. Indeed, the product
\begin{equation}
    \sigmaeff^{-1}\langle r^2 \rangle = \frac{(a_1+a_2)^2+4a_1a_2}{16\pi a_1a_2} 
\end{equation}
can be arbitrarily large. We can express it using $x=a_2/a_1$ as,
\begin{equation}
    \langle r^2\rangle= \frac{1}{16\pi}\left(x+6+\frac{1}{x}\right)\sigmaeff
\end{equation}
Similarly, the expected separation,
\begin{equation}
    \langle r \rangle = \frac{\sqrt{\pi}}{2\sqrt{2}} \left(\sqrt{a_1} + \sqrt{a_2}\right)
\end{equation}
can be expressed as,
\begin{equation}
    \langle r \rangle^2 = \frac{1}{128} \left(u^2+2u+4+\frac{8}{u}\right) \sigmaeff,
\end{equation}
where
\begin{equation}
    u=\sqrt{a_2 / a_1} + \sqrt{a_1 / a_2}. 
\end{equation}
Taking $a_1 / a_2 \to \infty$ or $a_2 / a_1 \to \infty$, and thus $u \to \infty$, shows that the expected separation $\langle r \rangle$ cannot be bounded by $\sqrt{\sigmaeff}$.

\begin{figure}[t]
    \centering
    \includegraphics[width=0.9\textwidth]{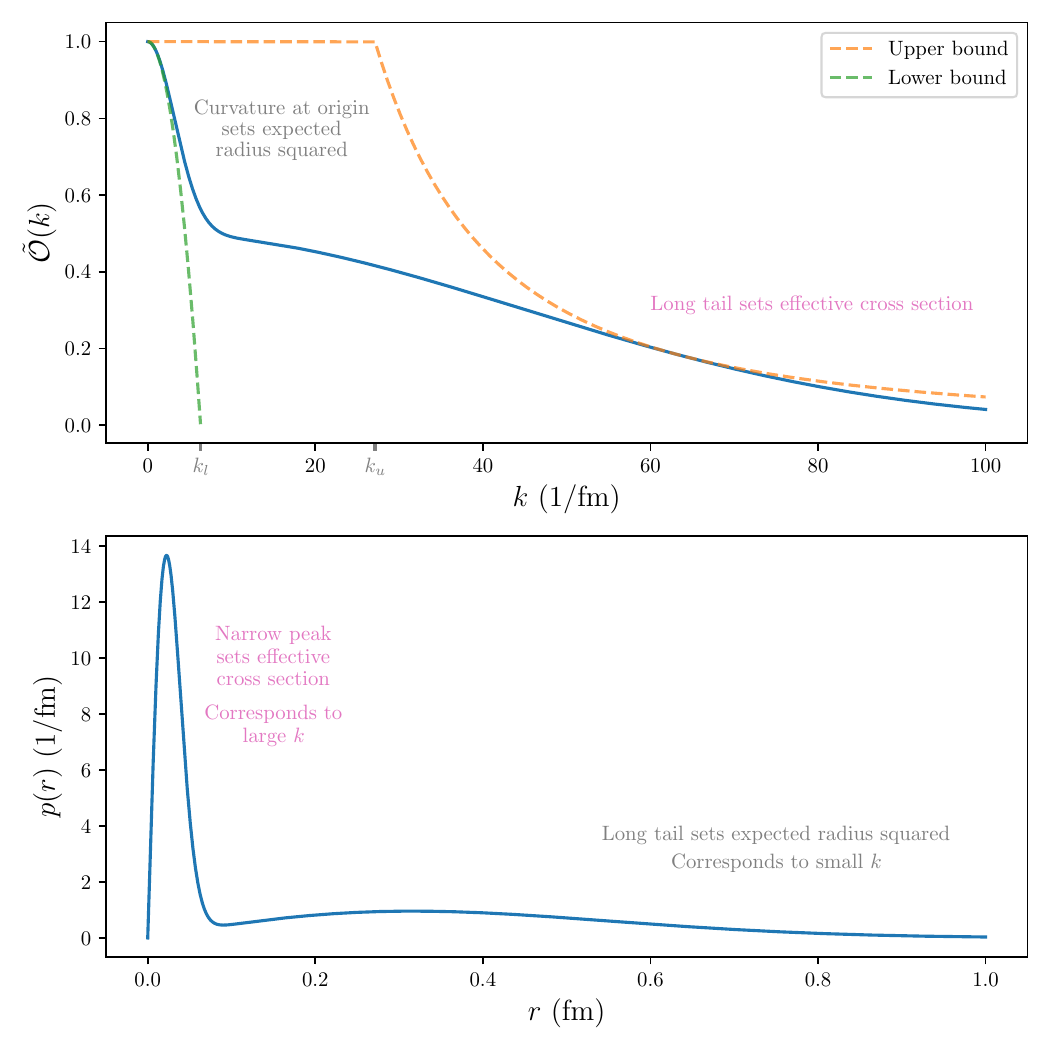}
    \caption{The structure function (upper) and radial density function (lower) for the example \cref{eq:counter_example} with $a_2 / a_1 = 200$, where the structure function is a mixture of two Gaussians. For reference, we show the assumed lower and upper bounds on the structure function that are used to construct our upper bound on the partonic separation, \cref{eq:assumption_low_k,eq:assumption_high_k} respectively.}
    \label{fig:counter}
\end{figure}

The inverse Fourier transform of a Gaussian is a Gaussian. Thus, in position space the radial pdf corresponding to \cref{eq:counter_example} equals,
\begin{equation}
    \pdfOneDim(r) = \frac{r}{2} \left(\frac{e^{-\frac12 r^2 / a_1}}{a_1} + \frac{e^{-\frac12 r^2 / a_2}}{a_2}\right) 
\end{equation}
Thus, \cref{eq:counter_example} is a simple, non-pathological counterexample to the claimed upper bound in~\red{Ref.}~\cite{Rinaldi:2018bsf}. The mixture of Gaussians in \cref{eq:counter_example} is a positive and monotonically decreasing function in momentum space and corresponds to a unimodal distribution on the transverse plane; there are no obvious additional assumptions that could exclude this case. We plot this example in radial position and momentum space in \cref{fig:counter}.

Dropping the assumptions in \cref{eq:assumption_low_k,eq:assumption_intermediate_k,eq:assumption_high_k}, one can construct even simpler examples in which the partonic separation is unbounded, e.g., an overlap function that is uniform between two concentric circles,
\begin{equation}\label{eq:simple_example}
    \overlap(x, y) = \begin{cases}
        \frac{1}{\pi(R_2^2 - R_1^2)} & \text{if $R_1^2 \le x^2 + y^2 \le R_2^2$}\\
        0 & \text{elsewhere.}
    \end{cases}
\end{equation}
One can readily obtain,
\begin{equation}
    \sigmaeff^{-1} = \frac{1}{\pi(R_2^2 - R_1^2)},
\end{equation}
whereas,
\begin{equation}
    \langle r^2 \rangle = \frac12 (R_1^2 + R_2^2).
\end{equation}
Thus, one can make the partons arbitrarily peaked ($R_1 \to R_2$) yet arbitrarily separated ($R_2 \to \infty$). In momentum space, it corresponds to
\begin{equation}
    \fourier{\overlap}(k) = \frac{2}{k(R_2^2 - R_1^2)} \left[R_2 J_1(R_2 k) - R_1 J_1(R_1 k)\right]
\end{equation}
The Bessel functions decay only as $1/\sqrt{k}$ and thus this structure function cannot satisfy the requirement \cref{eq:assumption_high_k}. 

\section{Conclusions}\label{sec:concs}

In this letter, we have investigated the extent to which double parton scattering (DPS) in proton collisions encodes information about the transverse separation between partons in the proton. We worked in the framework of the pocket formula \red{in order to re-examine and clarify previous results found in this framework. Assuming} only that the root-mean-square partonic separation existed, we showed that measurements of DPS effective cross sections imply the lower bound $\langle r^2\rangle \ge 4\sigmaeff/(9\pi)$ on the mean-squared partonic separation in the proton. A typical value of $\sigmaeff=15\mb$ gives $\rms{r} \ge 0.46\fm$. Whilst the existence of a lower bound was previously known, we improved it by a factor of $4/3$ and showed that it is the sharpest lower bound achievable under our assumptions. 

Contrary to previous works, we showed that in the framework of the pocket formula, $\sigmaeff$ cannot provide an upper bound on partonic separation without model-dependent assumptions on the transverse profile. The DPS effective cross sections depend only on the peakedness of the overlap function; even if measurements show that the overlap function is peaked, it need not be peaked at small separations. In particular, we showed through analysis and counterexamples that assuming rotational symmetry and that the structure function decays at least as $1/k^2$ in momentum space cannot lead to an upper bound; we present simple, non-pathological examples for which the partonic separation is unbounded, despite a finite effective cross section. \red{Although the pocket formula framework fails to model QCD effects that correlate longitudinal momenta and transverse separation, it serves as a toy model to build intuition and understanding. Using this toy model, our results correct a misconception about the relationship between the effective cross section and the proton's transverse structure.}

In summary, by treating the issue as a problem in mathematical statistics --- bounding the moments of a distribution given a constraint on the peakedness --- we shed light on the relationship between DPS measurements and partonic separation. These findings correct previous results in the literature and clarify what we can and cannot learn about proton structure from DPS measurements in the pocket formula framework.

\section*{Acknowledgements}

We thank Jan Kretzschmar for discussions and support. AF was supported by the National Natural Science Foundation of China (NNSFC) RFIS-II W2432006. HY was supported by XJTLU Postgraduate Research Scholarship PGRS2212033. 

\bibliographystyle{apsrev4-2j}
\bibliography{refs}

\end{document}